\documentclass[12pt]{article}
\pdfoutput=1
\usepackage{graphicx}
\usepackage{subfigure,amssymb,amsmath}
\usepackage{cite,color,xcolor,url,cancel,ulem}
\usepackage{bm}
\usepackage{float}
\usepackage{jheppub}
\def\be{\begin{equation}}
\def\bea{\begin{equation}}
\def\beqn{\begin{eqnarray}}
\def\ee{\end{equation}}
\def\eea{\end{equation}}
\def\eeqn{\end{eqnarray}}

\def\issue(#1,#2,#3){{\bf #1}, #2 (#3)}
\def\PREP(#1,#2,#3){Phys.\ Rep. \issue(#1,#2,#3)}
\newcommand{\ms}[2]{m_{\tilde{#1}_{#2}}}

\def\bea{\begin{equation}}
\def\eea{\end{equation}}
\def\bd{\begin{displaymath}}
\def\ed{\end{displaymath}}

\def\321{$\rm SU(3)_C\times SU(2)_L\times U(1)_Y$}

\def\vev#1{\left\langle #1\right\rangle}%

\newcommand{\wt}{\widetilde}

\def\10{SO(10)}

\def\sb{\tilde{b}}

\def\stau{\tilde{\tau}}
\def\stau#1{{\tilde \tau}_{#1}}
\def\smuon#1{{\tilde \mu}_{#1}}

\def\lspone{\widetilde\chi_1^0}
\def\mlspone{m_{\lspone}}

\def\ch2p{{\wt\chi_2^+}}

\def\ch2m{{\wt\chi_2^-}}

\def\chonepm{{\wt\chi_1}^{\pm}}

\def\mchonepm{m_{\chonepm}}

\def\bsg{{\rm Br}(B \rightarrow X_s +\gamma)}

\def\gmin2{{(g-2)}_\mu}
\def\amususy{a_\mu^{\rm SUSY}}

\def\issue(#1,#2,#3){{\bf #1}, #2 (#3)}%AIP format!Vol,page(Year)
\def\PREP(#1,#2,#3){Phys.\ Rep. \issue(#1,#2,#3)}
\newcommand{\mstopone}{m_{\tilde{t}_1}}

\def\lspone{\widetilde\chi_1^0}
\def\mlspone{m_{\lspone}}

\def\mul2{m_{{\tilde u}_L}^2}
\def\mur2{m_{{\tilde u}_R}^2}
\def\mdl2{m_{{\tilde d}_L}^2}
\def\mdr2{m_{{\tilde d}_R}^2}
\def\mel2{m_{{\tilde e}_L}^2}
\def\mer2{m_{{\tilde e}_R}^2}
\def\mtl2{m_{{\tilde t}_L}^2}
\def\mtr2{m_{{\tilde t}_R}^2}
\def\mbl2{m_{{\tilde b}_L}^2}
\def\mbr2{m_{{\tilde b}_R}^2}
\def\mtaul2{m_{{\tilde \tau}_L}^2}
\def\mtaur2{m_{{\tilde \tau}_R}^2}

\definecolor{light-gray}{gray}{0.75}

\title{Exploring Non-Holomorphic Soft Terms in the Framework of Gauge Mediated Supersymmetry Breaking}

\author[a]{Utpal Chattopadhyay,}
\affiliation[a]{Department of Theoretical Physics, Indian Association for the Cultivation of Science,\\  
	2A \& B Raja S.C. Mullick Road, Jadavpur, 
	Kolkata 700 032, India}

\author[b]{Debottam Das,}
\affiliation[b]{Institute of Physics, Bhubaneswar 751005,
India, and \\
Homi Bhabha National Institute,
Training School Complex, Anushakti Nagar, Mumbai 400085, India}
\author[a]{Samadrita Mukherjee}

\emailAdd{tpuc@iacs.res.in}
\emailAdd{debottam@iopb.res.in}
\emailAdd{tpsm9@iacs.res.in}

\abstract{
It is known that in the absence of a gauge singlet field, a specific class of 
supersymmetry (SUSY) breaking non-holomorphic (NH) terms can be soft 
breaking in nature so that they may be considered along with the
Minimal Supersymmetric Standard Model (MSSM) and beyond.
There have been
studies related to these terms in minimal supergravity based models.
Consideration of an F-type SUSY breaking scenario 
in the hidden sector with two chiral superfields 
however showed Planck scale suppression of such terms.
In an unbiased point of view for the sources of SUSY breaking, the NH terms  
in a phenomenological MSSM (pMSSM) type of analysis  
showed a possibility of a large SUSY contribution to muon $g-2$, a reasonable
amount of corrections to the Higgs boson mass and 
a drastic reduction of the electroweak fine-tuning for a 
higgsino dominated $\lspone$ in some regions of parameter space. 
  We first investigate here the effects of the NH terms in a low scale
SUSY breaking scenario. In our analysis with minimal gauge mediated
supersymmetry breaking (mGMSB) we probe how far the results can be compared 
with the previous pMSSM plus NH terms based study. 
We particularly analyze the Higgs, stop and the
electroweakino sectors focusing on a higgsino dominated
$\lspone$ and $\chonepm$, a feature typically different from what appears
in mGMSB. The effect of a limited degree of RG
evolutions and vanishing of the trilinear coupling terms
at the messenger scale can be overcome by choosing
a non-minimal GMSB scenario, such as one with a matter-messenger
interaction.
}
\begin{document}
\begin{flushright}
IP/BBSR/2017-12
\end{flushright}
\maketitle

\vspace{1cm}

\section{Introduction}
\label{sec:intro}
The discovery of Higgs boson with a mass of
125~GeV\cite{Aad:2015zhl,HiggsDiscoveryJuly2012} along with null
  search results for supersymmetry (SUSY) at the Large Hadron Collider (LHC)\cite{lhc1,lhc2} has in general enhanced the masses 
  of the superpartners of the low energy supersymmetry models.
  This is even more prominent in models like minimal supergravity (mSUGRA)\cite{msugra_orig} or
  Constrained Minimal Supersymmetric Standard Model (CMSSM)  
  where correlations among the masses via renormalization group evolution (RGE) effects make the spectra
  further heavy. 
  In the Minimal Supersymmetric Standard Model
  (MSSM)\cite{SUSYreviews1,SUSYbook1,SUSYbook2,SUSYreviews2}, 
unless the third generation scalars are themselves too massive,  
the Higgs mass data translates into the requirement of a
large top-squark left-right (L-R) mixing. This is required for obtaining the 
desired amount of radiative corrections for the mass of 
the CP-even lighter Higgs boson ($h$)\cite{djouadi}. This, on the other hand, 
increases the electroweak fine-tuning\cite{ewft_papers,baer_ewft,ewftotherworks,ewft_3,ewft_chan_chatto_nath}.

On the dark matter (DM) front, MSSM may provide with a suitable
candidate like the lightest neutralino, the lightest
supersymmetric particle (LSP)\cite{darkreview1,darkreview2}. The LSP in its 
bino dominated state needs very light sleptons to satisfy the
DM relic density limits from PLANCK experiment\cite{Ade:2015xua}. Light
sleptons is a rather disfavored scenario for most of the SUSY models in the
post-Higgs discovery era.  A bino can, however, be 
a favorable DM candidate, via its coannihilations with stau ($\stau1$) or
via its s-channel higgs exchange self-annihilation
mechanism\cite{mybinorefs,myfunnelrefs}.
In spite of the above, even with the possibility of having a reasonably light
LSP, the above coannihilation or resonance annihilations of binos require
stringent correlations among unrelated SUSY parameters. A mixed bino-higgsino
scenario like the Hyperbolic Branch\cite{ewft_chan_chatto_nath,HBnew}/Focus Point\cite{FP} zones can
still have a reasonably light LSP. However, it is not so favorable
via the spin-independent (SI) direct detection cross-section limits\cite{Draper:2013cka}.
The reason lies in the
larger values of Higgs-$\lspone$-$\lspone$ couplings. On the other hand,
in MSSM  
one can have a higgsino like LSP as a suitable dark matter candidate that
can pair-annihilate efficiently via their couplings to the W/Z gauge bosons
\cite{myhiggsinorefs}.  
Unlike a bino dominated LSP, a higgsino type of DM would not require any
fine-adjustment of uncorrelated SUSY parameters. 
%The SI direct detection cross-sections though low, can be probed in future
%XENON1t experiments.   
But a higgsino satisfying the 
observational relic density limits is typically quite
heavy\cite{myhiggsinorefs} ($\sim$~1 TeV)
\footnote{The limit goes below 600~GeV for higgsinos undergoing sfermion coannihilations\cite{Chakraborti:2017dpu}.} 
giving rise to a large electroweak
fine-tuning in MSSM. 
It is thus desirable to have a model that has a higgsino like LSP but is able
to guard fine-tuning to become large. We further note that a higgsino like
LSP scenario of phenomenological MSSM (pMSSM)\cite{pmssmref} is significantly free from the LHC
bounds\cite{WithADsecond}.

The other issue at hand in MSSM is the stringent muon $g-2$
constraint\cite{Bennett:2006fi,Gohn:2016ezs,pdgg-2}.   
Satisfying the muon $g-2$ constraint\cite{Davier:2017zfy} in the
MSSM is associated with a significant degree of reduction of
parameter space.   
As we know, SUSY contributions to muon $g-2 $ are enhanced 
when the lighter electroweakinos ($\chonepm$ and $\lspone$)
or the lighter smuon mass $\smuon1$ become small.
Concerning $\bsg$\cite{Amhis:2016xyh}, the constraint can be effective 
in MSSM for a high
$\tan\beta$\cite{Haisch:2012re,ourworkswherebsgismentionable}, the
ratio of Higgs vacuum expectation values 
and non-decoupled zone of squark masses.

All the above that restrict the MSSM parameter space may be 
addressed reasonably well when one includes non-holomorphic (NH) soft
terms on the top of 
the usual holomorphic soft terms of MSSM\cite{Un:2014afa2,Ca:nh}.
Staying within an MSSM setup terms like $\phi^2\phi^*$ or a
higgsino mass term like $\psi\psi$ can be soft SUSY breaking in nature if
there is no presence of a gauge singlet field\cite{Martin:nh}. 
There have been several works over the past two decades\cite{Hall:nh,Jack:nh,Jack:nh1,Frere:1999uv,Martin:nh,Hetherington:2001bk,Cakir:2005hd,Sabanci:2008qp,
  Solmaz2009,Un:2014afa2,Ross:2016pml,Ca:nh,Ross:2017kjc,Beuria:2017gtf} that analyzed the effects of introducing NH soft terms\footnote{See Ref.\cite{Haber:wh} for an analysis with 
  hard SUSY breaking terms.}. There are three distinct signatures i) the trilinear
NH terms may enhance L-R mixing thus influencing various phenomenologies, ii)
higgsinos get an additional soft-term source, iii) the Higgs potential remains unaffected at the tree level thus electroweak fine-tuning may hardly
change. The latter effect can give rise to a valid higgsino dark matter with
low fine-tuning, whereas the L-R mixing effect may contribute to $\gmin2$,
$\bsg$, or the higgs mass radiative corrections\cite{Sabanci:2008qp,
    Solmaz2009,Un:2014afa2,Ca:nh,Ross:2016pml,Ross:2017kjc}. The NH soft terms can also influence
on the scalar potential terms involving colored/charged scalars\cite{Beuria:2017gtf}.  Thus, there are 
associated charge and color breaking (CCB) minima 
constraints that one must take care of
in the limiting cases of large trilinear NH couplings.

All the above effects 
in combination may potentially cause very distinct outcomes in global
analyses of low energy SUSY models\cite{global_all}.  
In a model-dependent standpoint, one finds that such NH soft terms may 
arise out of interactions like 
$\frac{1}{M^3}{[XX^* \Phi^2 \Phi^*]}_D$ and 
$\frac{1}{M^3}{[XX^* D^\alpha \Phi D_\alpha \Phi]}_D$ \cite{Martin:nh}. 
In the above one considers
a scenario where supersymmetry is broken in the hidden sector
via an auxiliary field $F$ of the chiral
superfield $X$.  $\Phi$ is another chiral superfield and $M$ refers
to the mediation scale.  Both these terms have a strength
$\frac{{|F|}^2}{M^3} \sim \frac{M_W^2}{M}$\cite{Ca:nh}.
Clearly, there are large suppressions ($\sim 1/M$) for the NH soft terms in supergravity
scenarios where $M$ can be very large nearing the Planck scale. 
However, the suppression effects may be small if one considers a SUSY breaking mechanism that would allow a low mediation scale.  

With the above motivation, we like to explore the effects of having 
NH soft terms in gauge mediated supersymmetry breaking (GMSB) where 
SUSY breaking may occur at a smaller scale and the scale 
of mediation, the messenger scale is also low. We will try to gauge, how far the
electroweak scale analysis with NH soft terms of Ref.\cite{Ca:nh} referred as
Non-Holomorphic Supersymmetric Standard Model (NHSSM) preserves 
its conclusion in a 
minimal GMSB (mGMSB)
setup\cite {gmsb1,gmsb2,Giudice:1998bp,Giudice_Rattazzi_wavefn}. We remind
that the mGMSB
has gravitino as its dark matter candidate with its mass ranging from 
a few eV up to ${\cal O}$(1) GeV\cite{Giudice:1998bp,Dimopoulos:1996gy,Dimopoulos:1996yq,Baltz:2001rq,Fujii:2002fv,Jedamzik:2005ir,Feng:2010ij} whereas the
Next-to-lightest-SUSY Particle (NLSP) is largely a bino dominated
neutralino. We will focus here on realizing an mGMSB setup with  
higgsino as the NLSP.  This is 
unlike the typical case of a bino like NLSP that is characteristic of GMSB types of
analyses\cite {gmsb1,gmsb2,Giudice:1998bp}\footnote{Higgsino NSLP may be possible in an extended version of GMSB model that incorporates
  non-unified messenger sector and messenger-matter
  interactions\cite{Ahmed:2016lkh}.}.
We will investigate the effects of NH soft terms on
NLSP decaying to a gravitino and a Z-boson
$ \widetilde{\chi}^0_1 \to \widetilde{G}\ + Z$ 
or $ \widetilde{\chi}^0_1 \to \widetilde{G} +
h $\cite{Dimopoulos:1996yq, Giudice:1998bp} while assuming the other higgs
bosons to be much heavier than the NLSP.

We will now briefly describe the plan of the work. In
Sec.\ref{sec:modeldescription} we will outline the gauge mediated breaking
SUSY mechanism and introduce non-holomorphic soft terms within the
above framework. We will discuss the effect of NH terms on
Higgs and electroweakinos. 
In Sec.\ref{sec:results} we will describe the results of the relevant
parameter scanning on the top-squark and higgs boson masses and
investigate the phenomenologies involving $\bsg$ and
muon $g-2$. We will compare our results with the scenario of
MSSM with NH terms where one gives all the input parameters at the
electroweak scale. Furthermore, focusing on a parameter space where 
the NLSP is a higgsino dominated lightest
neutralino, a scenario typically unavailable in mGMSB, we will
estimate the relevant NLSP decay widths for a higgsino producing  
gravitino and other particles like $h$ and $Z$ bosons. Finally, we will
conclude in Sec.\ref{sec:conclusion}. 

\section{Gauge Mediated SUSY Breaking and
  Non-holomorphic Soft Terms} 
\label{sec:modeldescription}
In the Gauge-Mediated-SUSY breaking (GMSB) 
Models\cite{gmsb1,gmsb2, Giudice:1998bp} SUSY is broken in a
hidden/secluded sector. In the absence of the knowledge of the
hidden sector, one may consider a spurion $S$ which is a chiral
superfield, singlet under the SM group. SUSY breaking in this sector is realized via $S$ acquiring a vacuum expectation value (vev) 
via its scalar and auxiliary components. Thus,
\begin{equation}
\langle S \rangle=M + \theta \theta \langle F \rangle. 
\end{equation}  
The parameters $M$ and $\sqrt{\langle F \rangle}\equiv \sqrt F$ are the fundamental
scales related to GMSB. Apart from the hidden sector and the
observable sector where MSSM fields reside, there is a messenger sector
that itself experiences the SUSY breaking and mediates
the SUSY breaking it is affected with, to the observable sector. The
messenger sector connects to the hidden sector via the singlet spurion field 
$S$ that goes into a superpotential 
containing superfields of the messenger and
hidden sectors.
This results into a SUSY breaking via the two vevs of the
scalar and the auxiliary components of $S$ namely, 
$M$ and $\langle F \rangle$ causing 
a splitting of the masses of the messenger sector scalars
($m_{\phi,{\tilde \phi}}^2=M^2 \pm \langle F \rangle$ ) and fermions
($m_{\psi,\tilde \psi}=M$). The fact that the
scalars should not go tachyonic, so that the vacuum stability is unaffected
demands $\langle F \rangle < M^2$~\footnote{Realistic scenarios
  rather satisfy $\vev F<<M^2$.}. Considering a large $M$, we can integrate out the messenger scalars and fermions
that are charged under the SM gauge group
$SU(3)_C \times SU(2)_L \times U(1)_Y$.
The low energy effective theory below $M$ would then break supersymmetry
in the observable sector. 
 The observable sector soft terms
like the masses and couplings are generated via messenger 
particles in loops. This ensures renormalizability,
a positive feature of GMSB models. 
The MSSM soft terms thus obtained via gauge boson and
gaugino interactions are also flavor blind. 
It is the messenger sector particles, assumed to be
heavy for phenomenological reasons characterize the
low energy phenomenology. The gaugino masses arise out of one-loop diagrams such as that 
involving a messenger scalar and a fermion. The scalar mass squares arise out of two loop diagrams
that may include messenger scalars and fermions apart
from gauge bosons and gauginos.

Thus, the gaugino masses
$m_\lambda$ and scalar mass squares
$m_{\tilde f}^2$ read,
\begin{equation}
  m_\lambda \sim \frac{g^2}{16\pi^2} \Lambda
  \left(1+{\cal O}({\vev F}^2/M^4)\right); ~{\rm and,}~ \quad 
  m_{\tilde f}^2 \sim {\left(\frac{g^2}{16\pi^2}\right)}^2\Lambda^2\left(1+{\cal O}({\vev F}^2/M^4)\right),
  \label{gaugino_scalar_etc_eqn}
\end{equation}
where $\Lambda=\frac{\vev F}{M}$, is the SUSY breaking scale in MSSM.
The parameters $m_\lambda$ and $m_{\tilde f}$ come with similar
values. 
The trilinear parameters are small and
considered to be 
vanishing at the messenger scale, a result of the
fact that the messengers can only interact with SM fields
via gauge interactions. Vanishing of trilinear couplings
at the messenger scale may also be seen  
in Ref.\cite{Giudice_Rattazzi_wavefn} that used  
wavefunction renormalization method without a need of
using Feynman diagrams in this regard. 
The trilinear couplings associated with the third
generation of scalars are however
non-vanishing at the electroweak scale via
renormalization group evolutions. 
A soft mass value of 
$m_{\tilde f}\sim {\rm 1~TeV}$ would set
$\frac{\langle F \rangle}{M} \sim 10^5$~GeV
(Eq.\ref{gaugino_scalar_etc_eqn}), though this   
neither specifies $\vev F$ nor $M$. We note that the
scalars should not break vacuum stability, as mentioned
before, demands $\vev F<M^2$. In its limiting case
of an equality, one has $\frac{\vev F}{M} = M =10^5$~GeV. This results
into $M=10^5$~GeV and $\sqrt{\vev F} =10^5$~GeV. Considering the
inequality itself, one obtains the lower bounds
$M>10^5$~GeV and $\sqrt{\vev F} >10^5$~GeV.   

The upper limit of the messenger mass 
$M$ comes from the relative degree of strengths of the
gauge and gravity mediations since the superfield 
$S$ may also cause gravity effects,  
though in much smaller strength, in addition to the SUSY
breaking associated with a GMSB
scenario. The scalar mass squares from the gravity and
gauge mediated scenarios may now be compared. 
A representative factor of $1/1000$ for the ratio of the two
soft mass squares would mean
$\frac{g^2}{16\pi^2}\frac{\vev F}{M} > 10^{3/2} \frac{\vev F}{M_P} $,
where $M_P$ is the Planck mass\cite{Giudice:1998bp}. 
%\footnote{$M_P=  \sqrt{\frac{\hbar c}{8\pi G}}=2.435\times 10^{18}$~GeV}.
The above results into an approximate upper bound $M < 10^{15}$~GeV
\footnote{This is via considering a value of $g$ corresponding to
  the strong coupling at the electroweak scale.}.
Correspondingly, for a 1~TeV scalar mass, one finds 
$\sqrt{\vev F}<10^{10}$~GeV. Summarizing, one finds, 
\begin{equation}
10^5<\sqrt {\vev F}<10^{10}~{\rm GeV,~and,~}
10^5<M<10^{15}~{\rm GeV}.
\label{F_rangeEtc}
\end{equation}
Considering the supergravity relation 
the gravitino mass is given by $m_{3/2}=\frac{F}{\sqrt 3 M_P}$.
Depending on $\vev F$ (Eq.\ref{F_rangeEtc}), 
the gravitino mass may range from $\sim$1~eV to $\sim$1~GeV.    

Now, within the MSSM framework the NH soft terms are given by\cite{Un:2014afa2,Ca:nh}\footnote{There can be an NH soft term like $\psi \lambda$
  involving higgsinos and gauginos. This would however take us away from
  MSSM, hence ignored\cite{Martin:nh}.}:   
\begin{equation}
\label{nh_lagrangian}
 -\mathcal{L'}_{soft}^{NH} \supset \tilde{Q}\cdot H_{d}^c A_{u}'\tilde{U} + 
 \tilde{Q}\cdot H_{u}^c A_{d}'\tilde{D} +\tilde{L}\cdot H_{u}^c A_{e}'\tilde{E} +  
 \mu '\tilde{H_u}\cdot \tilde{H_d} +h.c.
\end{equation}
Here, $A_i'$ refers to three  
  $3\times 3$ matrices ($[A_i']$) belonging to
  the family space of squarks/sleptons.
  We remind that the MSSM soft terms are generated in mGMSB via
  gauge and gaugino
  interactions that are flavor blind. The messenger sector particles
  occur only in loops. 
  All the elements of 
  $[A_i]$, the associated parameters
  for holomorphic trilinear interactions that arise at two-loop levels
  are considered to be zero at the messenger
  scale\cite{SUSYreviews2,gmsb1,gmsb2, Giudice:1998bp}. 
  We note that Eq.\ref{nh_lagrangian} differs from the holomorphic trilinear
  soft SUSY breaking Lagrangian by a change in the Higgs fields. Thus, 
  $H_{u,d}$ are replaced by their conjugates in Eq.\ref{nh_lagrangian}
  with appropriate book-keeping for the associated scalar fields for 
  hypercharge assignments. 
%  With no other essential change than the fields  
%  $H_{u,d}$ being replaced by their conjugates as in Eq.\ref{nh_lagrangian}
%  while going from holomorphic to non-holomorphic trilinear interactions,
%  $[A_i']$ also behaves
  %  similarly as $[A_i]$.
  Hence, as with the elements of $[A_i]$, the same for
  $[A_i']$ should also vanish at the
  same scale.
  With a limited zone of RG running typical to a GMSB scenario both $[A_i]$
  and $[A_i']$ terms are tiny at the electroweak scale for the first two
  generations. Thus $[A_i']$ terms do not cause any violation of
  flavor changing neutral current (FCNC) constraints that are typically
  stringent for observables related to the first two generations of fermions. 
%  With no structural change from 
%  mGMSB, here presence of $[A_i']$ terms would also be safe in relation to 
% flavor changing neutral current (FCNC) constraints.
  %The smallness of the RG evolutions of the trilinear couplings
  %for the first two generations 
  %is visible in Table~\ref{bptable}. Unlike mGMSB, these would not be  
  %so small in $N=1$ supergravity based models with a large range of energy
  %scale to run across.

As mentioned in 
Sec.\ref{sec:intro}, terms like $\phi^2\phi^*$ and a higgsino mass
soft-term $\psi \psi$ (Eq.\ref{nh_lagrangian}) 
may originate from D-term
contributions like 
$\frac{1}{M^3}{[XX^* \Phi^2 \Phi^*]}_D$ and
$\frac{1}{M^3}{[XX^* D^\alpha \Phi D_\alpha \Phi]}_D$\cite{Ca:nh,Martin:nh}. 
Since the terms are of strength $~\frac{M_W^2}{M}$, a low mediation scale $M$ such as
that appears in a GMSB scenario may be relevant for probing the
phenomenological implications. The parameters for trilinear non-holomorphic
soft terms are small, similar to the same of trilinear holomorphic terms at the mediation scale $M$. 
We will keep the other D-term soft breaking NH interaction namely
the higgsino mass soft term characterized by $\mu^\prime$ (to be given
at the scale $M$), to have an unknown SUSY
breaking origin\footnote{Additionally, such a term cannot originate
  at a one-loop level in
  mGMSB, thus becoming suppressed.}. This is considering the issue of an associated
re-parametrization invariance of the higgsino mass
soft term\cite{Ross:2016pml,Ross:2017kjc,Hetherington:2001bk,Jack:nh}.
Reparametrization comes from unrelated quantities like $\mu$, the higgsino
mixing superpotential parameter and the higgs scalar
soft mass parameters $m_{H_U}^2$ and $m_{H_D}^2$.  An assumption of an 
independent SUSY breaking mechanism to have a higgsino mass soft term
essentially
avoids such concerns (See discussions in \cite{Ross:2016pml,Ross:2017kjc} along with the references therein).

Coming to the minimal GMSB, to preserve the
gauge coupling unification one considers
messengers to belong to a 
complete $SU(5)$ representation or any other complete representation of a 
larger gauge group that includes $SU(5)$ as a sub-group.
With $S$ as the spurion field mentioned before one has the superpotential, \begin{equation}
W_{\rm mess}=S \Phi {\bar \Phi}.
\end{equation}    
In the simplest case, there will be $N_5$ number of flavor of
messenger copies $\Phi$ and
$\bar \Phi$ transforming as $5$ and $\bar 5$ representations of SU(5).
The soft SUSY
breaking parameters like gaugino and scalar masses are given as follows.
\begin{equation}
 \label{gaugino_mass}
M_\alpha = \frac{g_\alpha^2}{16\pi^2} N_5 \Lambda g(x), \quad(\alpha=1,2,3). \end{equation}
Here $\alpha$ refers to SM gauge group (U$(1)_Y$, SU$(2)_L$, \& SU$(3)_c$). \\
 $x=\frac {\langle F\rangle}{M^2}$,
$\Lambda=\frac {\langle F\rangle}{M}$ and $g(x)$ is given as,
\begin{equation}
  g(x) =\frac{1}{x^2}\left(
(1+x){\rm log}(1+x)+(1-x){\rm log}(1-x)
  \right).
\end{equation}
The scalar masses are given by,
\begin{equation}
\label{scalar_mass}
m_{\tilde f}^{2} = 2\Lambda^2 N_5 \sum_{\alpha}
\left(\frac{g_\alpha^2}{16\pi^2}\right)^2 C_\alpha  f(x). 
 \end{equation}
Here, $f(x)$ is
given by,
\begin{equation}
  f(x)=\frac{1+x}{x^2}\left(\log(1+x)-2\mathrm{Li}_2\left(\frac{x}{1+x}
  \right)+\frac{1}{2}\mathrm{Li}_2\left(\frac{2x}{1+x}\right)\right)+
  (x\rightarrow -x).
\end{equation}
$C_{\alpha}$ is the quadratic Casimir of the representation of the gauge group factor $G_{\alpha}$ under which the scalar field
$\tilde f$ transforms. 
For the gauge groups involved, the Casimirs are:
$ C_{SU(n)} = \frac{n^2-1}{2n}$ and $C_{U(1)} = (3/5)Y^2$.
The Euler dilogarithm function $\mathrm{Li}_2(x)$ is given by
$\mathrm{Li}_2(x)=-\int_0^x \frac{{\rm log}(1-t)}{t}dt$, for
$0<x<1$.
Under the assumption of $\langle F \rangle <<M^2$ one has $g(x) \simeq 1$ and
$f(x) \simeq 1$. This results into,
\begin{equation}
 \label{gaugino_mass1}
M_\alpha = \frac{g_\alpha^2}{16\pi^2} \Lambda N_5, 
\end{equation}
and,
\begin{equation}
\label{scalar_mass1}
m_{\tilde f}^{2} = 2\Lambda^2 N_5 \sum_{\alpha}
\left(\frac{g_\alpha^2}{16\pi^2}\right)^2 C_\alpha. 
 \end{equation}
%Furthermore, as discussed before the trilinear couplings are considered
%to be vanishing at the messenger scale $M$. 
α
The following are the model parameters where the   
NH higgsino soft parameter $\mu_0^\prime$ is the input value of $\mu^\prime$
at the messenger scale $M$ at which the common
NH trilinear coupling parameter $A_0^\prime$ vanish. 
\begin{equation}
\Lambda, M, \tan\beta, N_5, sign(\mu), {\rm and}~ \mu^{\prime}_{0}.
\end{equation}

\noindent

The NH trilinear couplings of Eq.\ref{nh_lagrangian} 
modify the off-diagonal elements of the scalar mass matrix as given
below\cite{Un:2014afa2,Ca:nh}. 
\begin{eqnarray}
\label{stop_mass}
 M_{\tilde{u}}^2=&\left(\begin{matrix}
   m_{\tilde{Q}}^2+(\frac{1}{2}-\frac{2}{3}\sin^2\theta_{W})M_Z^2\cos2\beta +m_u^2&\hspace{2mm}
   -m_u(A_u-(\mu+A_u')\cot\beta) \\
   -m_u(A_u-(\mu+A_u')\cot\beta)  & \hspace{-2mm}
   m_{\tilde{u}}^2+\frac{2}{3}\sin^2\theta_{W} M_Z^2 \cos2\beta +m_u^2 
                    \end{matrix} \right).\hspace{6mm}
%\label{stopmassmatrix}
\end{eqnarray}
Similar mass matrices for sleptons (or a down-type of squarks) would have the off-diagonal term
$-m_e(A_e-(\mu+A_e')\tan\beta)$.
Clearly, the NH trilinear couplings contribute to the off-diagonal elements where $\mu$ is 
replaced by $\mu+A_f'$ (${f = u,d,e ~ {\rm etc}.}$).  The above indicates
a more significant impact of L-R mixing
for i) low values of $\tan\beta$ for the up type of squarks and ii) high values of $\tan\beta$ for down type of squarks or sleptons. 

Now, the discovery of the  Higgs boson with mass of
$125.09 \pm 0.24$ GeV \cite{Aad:2015zhl,HiggsDiscoveryJuly2012}
is translated into  
a large amount of radiative corrections to the mass of the lighter
neutral CP-even Higgs boson $h$.  
The above requirement causes an increase in the masses of the top-squarks
in MSSM and/or one needs an appropriately large
value of $|A_t|$, so as to have a
stronger $\tilde t_L-\tilde t_R$ mixing.  Thus, the NH soft trilinear
parameter $A_t'$ that affects 
the L-R mixing has important contributions toward the above 
radiative corrections for small values of $\tan\beta$.

\noindent
The lightest CP-even higgs boson mass up to one loop
can be read as follows\cite{djouadi}. 
\begin{equation}
\label{higgs_loop}
m_{h,top}^2= m_Z^2\cos^22\beta + \frac{3 g^2_2 {\bar m}_t^4}{8 \pi^2 M_W^2} 
\left[\ln\left(\frac{\ms{t}{1} \ms{t}{2}}{{\bar m}_t^2}\right) + \frac{X_t^{'2}}{\ms{t}{1}\ms{t}{2}} 
\left(1 - \frac{X_t^{'2}}{12\ms{t}{1}\ms{t}{2}} \right) \right]. 
\end{equation}
Here, $X_t'=A_t-(\mu+A_t')\cot\beta$.  Clearly, $A_t'=0$ corresponds to 
the MSSM result.  Here ${\bar m}_t$ refers to the running top-quark mass 
that includes corrections from the electroweak, quantum chromodynamics (QCD)
and SUSY QCD related effects\cite{Pierce:1996zz}.  The maximal mixing scenario
refers to 
$X^{\prime}_t={\sqrt 6} M_S$ where $M_S=\sqrt{\ms{t}{1}\ms{t}{2}}$. 
With a suitable $A_t'$, it is possible to satisfy 
the Higgs mass constraint with a relatively smaller value of $|A_t|$ in NHSSM, compared to MSSM.

The changes in the neutralino and chargino mass matrices are as shown below. 
Essentially, $\mu$ is replaced by $\mu+\mu^\prime$ as in the tree level results.
\begin{eqnarray}
\label{neutralino_mass}
 M_{\widetilde{\chi^0}}=&\left(\begin{matrix}
                        M_1 & 0 & -M_Z\cos\beta \sin\theta_W & M_Z\sin\beta \sin\theta_W \\
                        0 & M_2 & M_Z\cos\beta \cos\theta_W & -M_Z\sin\beta \cos\theta_W \\
                       -M_Z\cos\beta \sin\theta_W & M_Z\cos\beta \cos\theta_W & 0 & -(\mu+\mu^\prime)\\
                        M_Z\sin\beta \sin\theta_W & -M_Z\sin\beta \cos\theta_W & -(\mu+\mu^\prime) & 0
                                           \end{matrix} \right).\hspace{6mm}
\end{eqnarray}
%Similarly for the chargino matrix we have :
%\cite{Ross:2016pml}:
\begin{eqnarray}
\label{chargino_mass}
  M_{\widetilde{\chi^{\pm}}}=&\left(\begin{matrix}
                                 M_2 & \sqrt{2} M_W\sin\beta \\
                                 \sqrt{2} M_W\cos\beta & (\mu+\mu^\prime)\\
                                \end{matrix}\right).
\end{eqnarray}
We note that the collider bound on the lighter chargino mass, except the
issue of radiative corrections, will 
apply to essentially $|\mu+\mu^\prime|$
instead of $|\mu|$ in case $\chonepm$ is higgsino dominated in nature. 
Since the Higgs potential is unaffected,
the electroweak fine-tuning measure at tree level 
is still dependent on $\mu$\cite{Ross:2016pml,Ca:nh,Ross:2017kjc}
rather than it has anything to do with $\mu^\prime$. A large higgsino mass
with less electroweak fine-tuning becomes a possibility.

In contrast to the NHSSM that is based on electroweak scale inputs 
we should keep in mind that the dependence of relevant quantities
especially $\mu$ on $\mu^\prime$ and $A_t'$ can be significant
in NHmGMSB due to RGE effects.
Besides, it is important to mention that a fine-tuning measure analyzed 
in a scenario with NH terms in a predictive model that uses RGEs can be   
less independent with respect to the higgsino mass\cite{Ross:2017kjc}.  
This is unlike what is seen in electroweak fine-tuning in 
NHSSM as mentioned above.

\section{Results}
\label{sec:results}
We have realized the non-holomorphic MSSM on a GMSB setup, that is going
to be referred as the NHmGMSB model,
by using the codes SARAH-4.9.1\cite{sarah} and SPheno-3.3.8\cite{Spheno}.
Two-loop RGEs of the MSSM soft parameters, plus the same at one-loop 
for the NH soft parameters are used to generate the sparticle mass
spectra\cite{Ross:2016pml,Jack:nh,Jack:nh1,RGEs}.
The codes use two-loop corrections
for Higgs states\cite{2loopHiggs} and additionally compute all the
relevant flavor observables\cite{avelino}. 
As mentioned earlier, the free parameters that are to be scanned are 
\bea
\label{parameters}
\Lambda,M_{\rm mess}~\footnote{Henceforth we will refer $M_{mess}$ as the messenger scale.}, \tan\beta, N_5, {\rm sign}(\mu) \rm ~and~ \mu^{\prime}_0 ~~,
\eea 
where the number of the messenger copies is taken to be one ($N_5=1$).
At the messenger scale $M_{\rm mess}$, 
the soft SUSY breaking mass parameters are given 
via Eqs.\ref{gaugino_mass},\ref{scalar_mass} and   
the values of (non) holomorphic trilinear coupling parameters
$(A^{\prime}_0),A_0$ are taken to be zero.
Choosing ${\rm sign}(\mu)=1$ 
%along with $\tan\beta = 10~ {\rm and}~ 40$,   
we scan the following volume of the mGMSB parameters.
\begin{align}
\label{scan_range}
 3.0 \times 10^5~ {\rm GeV} &\leqslant \Lambda \leqslant 1.0 \times 10^6~ {\rm GeV}, \nonumber \\
 2 \times 10^6~{\rm GeV} &\leqslant M_{\rm mess} \leqslant
 10^8~ {\rm GeV}, \nonumber \\
 \tan\beta &= 10~ {\rm and}~  40,  \\
% A_0^{\prime} &= 0, \nonumber \\
% \mu & > 0, \nonumber \\
 -4000~ {\rm GeV} &\leqslant \mu^{\prime}_0 \leqslant 4000~ {\rm GeV}, \nonumber
\end{align}
with $m_t^{pole}=173.5$~GeV, $m_b^{\overline{MS}}=4.18$~GeV and $m_\tau=1.77$~GeV \cite{Olive:2016xmw}
and $M_{\rm SUSY} = \sqrt{\ms{t}{1} \ms{t}{2}}$. Additionally, we note that
a higher value of $M_{\rm mess}$ than what is given in Eq.\ref{scan_range}
would not be so consistent with our motivation of choosing a low scale
SUSY breaking model like mGMSB. This is in keeping with limiting the
mediation scale suppression of the NH soft terms as mentioned in
Sec.\ref{sec:intro}. On the other hand, reducing $M_{\rm mess}$ further
than the lower limit of Eq.\ref{scan_range} would hardly provide with
a reasonable range of RG running of both $A_t$ and $A_t'$
that is essential for satisfying the higgs mass data (without trying to
compensate it via choosing a larger value of $\Lambda$).  
 
\noindent
The SUSY Higgs mass ($m_h$) limits\cite{Olive:2016xmw} and the 
constraints from B-physics namely $B \rightarrow X_s +\gamma$, 
$B_s \rightarrow \mu^+ \mu^-$ are mentioned as 
below\cite{Olive:2016xmw}. 
\begin{align}
\label{experimental_bounds}
122.1 ~ {\rm GeV} &\leqslant m_h \leqslant 128.1 ~{\rm GeV},  \nonumber \\
%m_{\chi^{\pm}} &\geqslant 104~ {\rm GeV}, \nonumber \\
2.99 \times 10^{-4}~&\leqslant Br(B \to X_s + \gamma) \leqslant 3.87 \times 10^{-4}~ (2\sigma), \\
1.5 \times 10^{-9} ~ &\leqslant Br(B_s \to \mu^+\mu^-) \leqslant 4.3 \times 10^{-9}~(2\sigma). \nonumber 
\end{align}
We consider a 3~GeV theoretical uncertainty in computing $m_h$ as given above.
Some of the reasons behind considering the spread are uncertainty in
computing loop corrections up to three loops, top quark mass, renormalization
scheme and scale dependence etc.\cite{loopcorrection}. 

We will now discuss the effect of introducing NH parameters within the mGMSB
scenario and especially study the dependence of the
SUSY spectra and observables  
of phenomenological interest on $\Lambda$, $M_{\rm mess}$ and $\mu^{\prime}_0$. We remind that in this analysis the trilinear NH couplings are vanishing at the messenger
scale whereas the higgsino mass soft-parameter $\mu_0^\prime$ is considered to arise from sources other than GMSB. This is following the discussion of Sec \ref{sec:modeldescription}.

\subsection{Effects on top-squarks and Higgs masses}

\begin{figure}[!ht]
    \begin{center}
        \subfigure[]{%       
          \includegraphics[width=0.5\textwidth,angle=0]{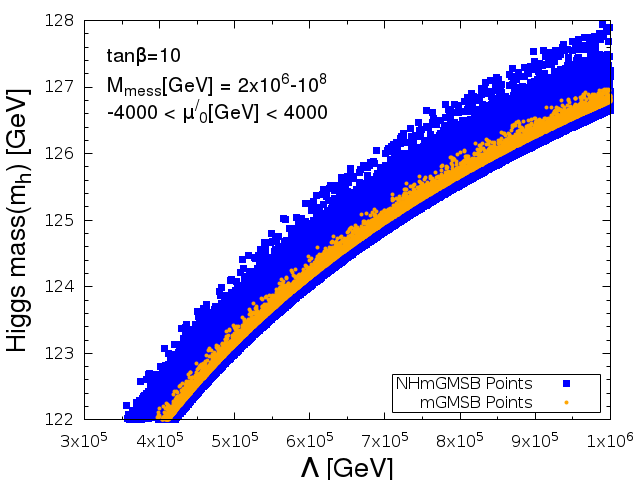}
          \label{fig:lambda-higgsA}
            }%
            \subfigure{%
              \includegraphics[width=0.5\textwidth,angle=0]{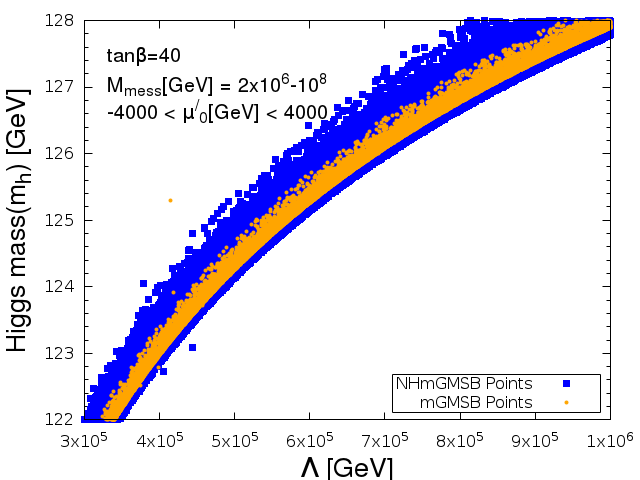}
              \label{fig:lambda-higgsB}
            }
            \caption{
              Scatter plots of $m_h$ with  $\Lambda$
              are shown for $\tan\beta=10$ and $40$ when
              $M_{mess}$ and $\mu_0^{\prime}$ are scanned
              according to the ranges mentioned in Eq.\ref{scan_range}. 
 The blue and yellow coloured regions correspond to NHmGMSB and mGMSB
    cases respectively.}
    \label{fig:lambda-higgs}
\end{center}
\end{figure}
Figs. \ref{fig:lambda-higgsA} and \ref{fig:lambda-higgsB}
represent the scatter plots of Higgs boson mass with
$\Lambda$ for $\tan\beta=10$ and $40$ respectively when
              $M_{mess}$ and $\mu_0^{\prime}$ are scanned
              according to Eq.\ref{scan_range}.
Here, the mass of the lighter CP-even neutral Higgs boson $h$
which is standard-model like in its couplings satisfies the
range mentioned in Eq.\ref{experimental_bounds}.
The yellow and blue zone refer to the results of mGMSB and NHmGMSB respectively.
The lower limit of $\Lambda$ corresponding to the lower
limit of $m_h$ (Eq.\ref{experimental_bounds}) for any given scenario either mGMSB or NHmGMSB
decreases with an increase in $\tan\beta$. 
The spread of points for $m_h$ shown in blue (NHmGMSB) is quite large
compared to the yellow region (mGMSB) for $\tan\beta=10$. However, 
this is not so for the case of $\tan\beta=40$. 
We note that $\mu+A_t'$ appearing in the off-diagonal
elements of the top-squark mass matrix is suppressed by $\tan\beta$. 
The larger top-squark mixing causes more prominent spread for low $\tan\beta$.
Regarding $A_t$, it is seen that 
$A_t$ turns out to be negative. However, $X_t'$ is not large enough to
be in the maximal mixing zone characterized by
$X_t'={\sqrt 6} M_S$. This is also true for 
$X_t$ ($=A_t-\mu\cot\beta$) for the mGMSB case.
Furthermore, it is found that $A^\prime_t$ at the EWSB scale approximately
varies from $-550~(-600)$~GeV 
to $550~(600)$~GeV for $\tan\beta = 10~(40)$.
$A^\prime_t$ comes with either sign because of its dependence via RGE on 
$\mu_0^\prime$, while the latter is scanned for both positive and negative
values.  Additionally, we note that $\mu_0^\prime$ may also contribute\cite{Ross:2016pml,Jack:nh,Jack:nh1}
quite significantly to the value of $\mu$ which is obtained
via REWSB. With all the above effects and estimates, $X_t'$ 
is seen to be larger in the negative direction for $\tan\beta=10$ than
$\tan\beta=40$.
Because of a larger top-squark mixing, that results into a larger radiative
corrections to the Higgs boson mass, the lowest value of 
$\Lambda$ satisfying the lower limit of $m_h$ for NHmGMSB is less than the same 
for mGMSB.

\begin{figure}[!htbp]
 \begin{center}
 \subfigure[]{%
  \includegraphics[width=0.5\textwidth,angle=0]{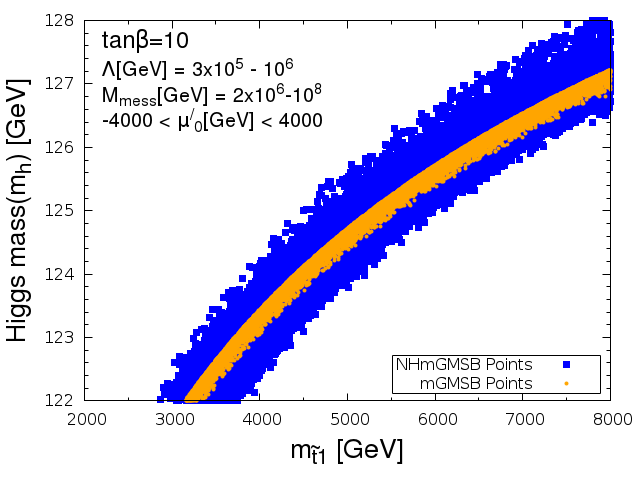}
\label{fig:higgs-stopA}
 }%
 \subfigure[]{%
   \label{fig:higgs-stopB}
  \includegraphics[width=0.5\textwidth,angle=0]{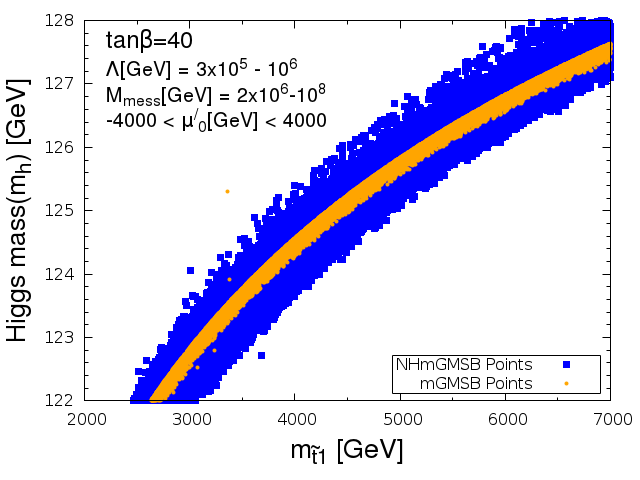}
  }
  \caption{\it Scatter plots of $m_h$ with stop mass $\mstopone$ are 
    shown for $\tan\beta=10$ and $40$. 
    $\Lambda$, $M_{mess}$, $\mu_0^{\prime}$ are scanned according to the ranges
    mentioned in Eq.\ref{scan_range}. The colors carry the same
    convention as that of Fig.\ref{fig:lambda-higgs}.
}  
 \label{fig:higgs-stop}
\end{center}
\end{figure}
We  will now explore the variation of Higgs boson mass $m_h$ with top-squark mass
$\mstopone$ for $\tan\beta=10$ and $40$ in Figs.\ref{fig:higgs-stopA} and Fig.\ref{fig:higgs-stopB} respectively,   
when $\Lambda$, $M_{\rm mess}$, $\mu^{\prime}_0$ are scanned according to
Eq.\ref{scan_range}. The colors carry the same convention as that of
Fig.\ref{fig:lambda-higgs}.
For a given value of $m_h$, the lower limit of $\mstopone$ becomes smaller
in NHmGMSB than that of mGMSB and this is more prominent for $\tan\beta=10$
because of a larger influence of $(\mu+A_t')$.
   For a given $\mstopone$, the amount of corrections in $m_h$ via NH terms
may reach up to nearly 1~GeV for $\tan\beta=10$ and 0.5 GeV for $\tan\beta=40$.
On the other hand, for a given $m_h$, the spread of $\mstopone$ is
about 1~TeV from its minimum to the maximum value for NHmGMSB within which
the scattered points for mGMSB reside.

\begin{figure}[!htbp]
 \begin{center}
    \subfigure[]{%
     \includegraphics[width=0.5\textwidth,angle=0]{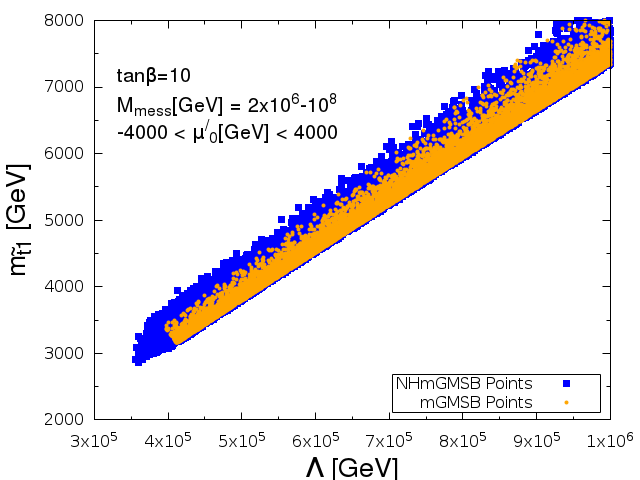}
\label{fig:lambdastopA}
    }%
    \subfigure[]{%
    \includegraphics[width=0.5\textwidth,angle=0]{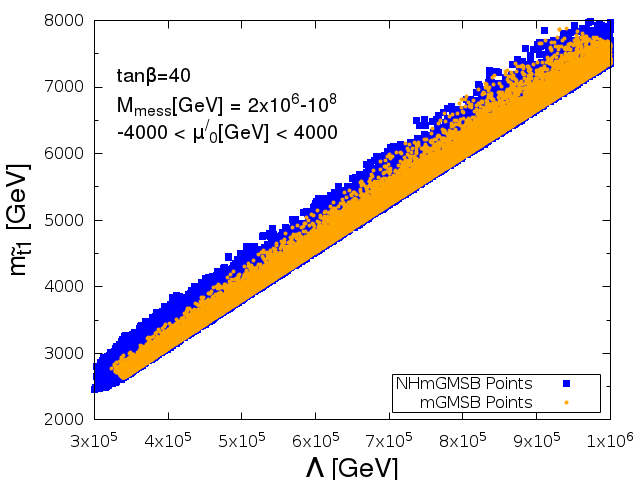}
\label{fig:lambdastopB}
    }
    \caption{Scatter plots of the lighter top-squark mass $\mstopone$
      with $\Lambda$ for $\tan\beta=10$ and $40$ for the scanning ranges 
      mentioned in Eq.\ref{scan_range}. The blue and yellow zones refer
      to the results of NHmGMSB and mGMSB respectively.}
 \label{fig:lambdastop}
 \end{center}
\end{figure}
Figs.\ref{fig:lambdastopA} and \ref{fig:lambdastopB} show
scatter plots of the lighter top-squark
mass $m_{{\tilde t}_1}$ with $\Lambda$  
corresponding to $\tan\beta=10$ and $40$ respectively, 
where the Higgs boson mass ($m_h$) satisfies the
range of Eq.\ref{experimental_bounds}.
The blue and yellow zones refer to the results of NHmGMSB and mGMSB
respectively. The Higgs boson mass lower limit
is reached for a smaller value of $\Lambda$
for $\tan\beta=40$ in comparison with $\tan\beta=10$ for reasons described earlier. 
With the validity of the lower $\Lambda$ zone for a larger $\tan\beta$ via
the $m_h$ lower limit, 
the top-squark mass $\mstopone$ finds its smallest value to be 
smaller for $\tan\beta=40$ compared to the same for $\tan\beta=10$. 
   On the other hand, between $\tan\beta=10$ and $40$, 
the relative difference of the lowest value of $\mstopone$ between NHmGMSB
and mGMSB is more enhanced for the lower value of $\tan\beta$ primarily
because of a lesser degree of suppression of the contribution of
$\mu+A_t'$ in the top-squark mixing.  The resulting reduction of
$\mstopone$ in NHmGMSB (compared to
mGMSB) is about 500 ~(200)~GeV for $\tan\beta=10$~(40).

\begin{figure}[!htbp]
      \begin{center}
   \subfigure[]{%
     \includegraphics[width=0.5\textwidth,angle=0]{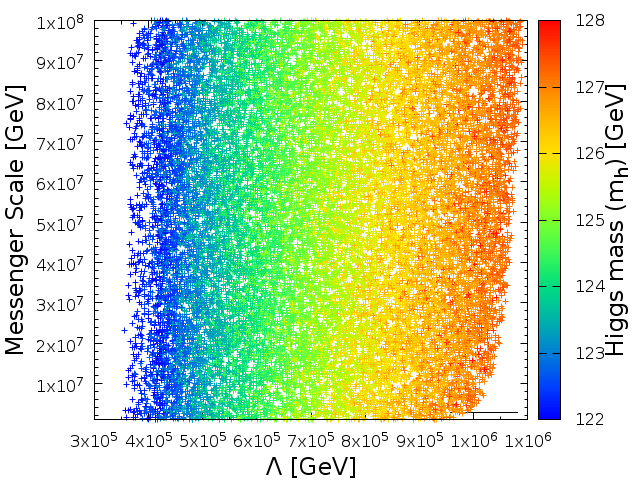}
        \label{fig:lambda-messengerA}
             }%
        \subfigure[]{%
          \includegraphics[width=0.5\textwidth,angle=0]{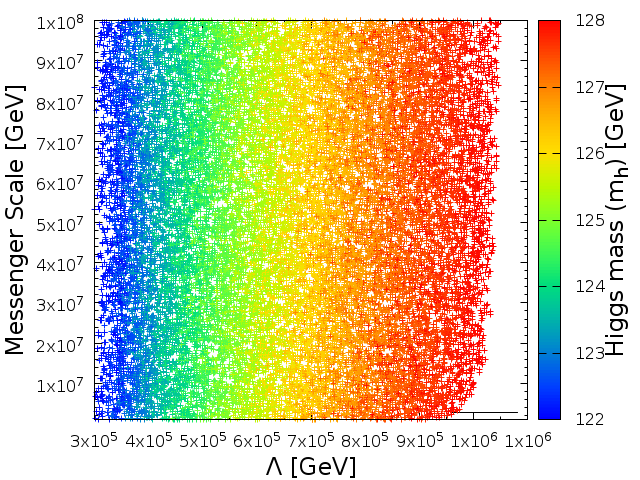}
        \label{fig:lambda-messengerB}}
        \caption{Scatter plots of $\Lambda$ against Messenger Scale
          $M_{mess}$ for $\tan\beta=10$ and $40$ for NHmGMSB, corresponding to
          the scanning ranges 
      mentioned in Eq.\ref{scan_range}. 
          The side panels of each
          of the plots show the Higgs mass. }
   \label{fig:lambda-messenger}
\end{center}
\end{figure}

We will now explore the role of the messenger scale $M_{\rm mess}$ in our
analysis that would shed some light on the  
extent of RG evolutions of relevant parameters that have influence 
on the scalar masses including that of the Higgs boson as well as 
on the masses of the electroweakinos.  The effects of evolutions
are not likely to be as large as that occur 
in minimal supergravity (mSUGRA)\cite{Un:2014afa2} with much higher SUSY
mediation scale,
but as we have seen they may assist in lowering  the top squark
mass $m_{{\tilde t}_1}$ or enhancing $m_h$.  
Fig.\ref{fig:lambda-messengerA} and \ref{fig:lambda-messengerB} show color contour plots in the 
($\Lambda$ vs $M_{mess}$)  plane for NHmGMSB corresponding to $\tan\beta=10$ and $40$ respectively when $\mu_0^\prime$ is
scanned as in Eq.\ref{scan_range}.  
The side panels show the values of $m_h$. The figures show strong dependence of Higgs mass on $\Lambda$ (that
sets the masses of the scalars as well as gauginos) and a weaker dependence on $M_{mess}$.
However, we must mention that the upper limit of $M_{mess}$ is chosen 
relatively small in our analysis. This is 
based on our motivation to have the NH terms of being associated with a lesser
degree of mass suppression as mentioned in Sec.\ref{sec:modeldescription}, the
reason of our working in a GMSB setup.   
A larger $M_{\rm mess} \sim 10^{12}$~GeV (not shown in this work) 
may increase both $A_t$ and $A_t'$ significantly to higher values
beyond what would be necessary to satisfy the Higgs boson mass limit 
for a reasonably chosen $\Lambda$. 

It is important to note that there are a few beyond 
the mGMSB analyses\cite{Ahmed:2016lkh,beyond_mGMSB} for example those 
involving matter-messenger interactions in which the  
trilinear holomorphic soft terms may arise at one-loop level, thus  
becoming non-vanishing at the messenger scale. In such situations one obtains   
a significantly large amount of radiative corrections
to $m_h$, a friendly feature to accommodate a Higgs boson as massive as 125 GeV
in models away from mGMSB.  The NH trilinear terms will then 
similarly become non-vanishing at the above scale. 
Depending on the sign of $\mu_0^\prime$, 
this may give rise to a larger $A_t'$ at the
electroweak scale. The above in turn that may result into a smaller $\mstopone$ or
positively contributing toward the radiative corrections to the Higgs
boson mass $m_h$.  Thus a smaller $\Lambda$ would be acceptable lowering the
overall mass scale of sparticles.
Additionally, as we will see below, the same consideration may
significantly enhance the smuon mass mixing via a larger $A_\mu'$. A large 
$A_\mu'$ may effectively contribute to the SUSY contribution to muon
$g-2$ significantly, because of an associated scaling by
$\tan\beta$\cite{Ca:nh}.   

\subsection{Phenomenological implications through the electroweakino sector}
\begin{figure}[!htbp]
 \begin{center}
  \includegraphics[width=0.5\textwidth,angle=0]{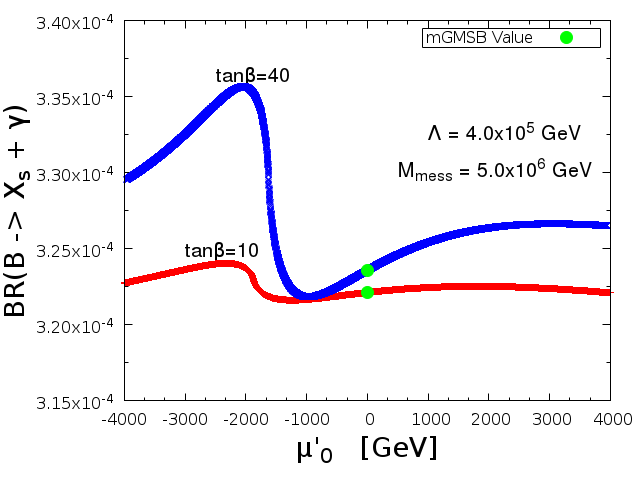}
    \caption{The variation of $Br(B \rightarrow X_s +\gamma)$ with $\mu^{\prime}_0$ 
      is shown for the shown values of $\Lambda$ and $M_{mess}$. The red and
      blue lines correspond to $\tan\beta=10$ and $40$
      respectively. Green filled circles on the top of the
      lines specify the corresponding mGMSB values.}
  \label{fig:b_sg}
 \end{center}
\end{figure}

As discussed in Sec.\ref{sec:modeldescription} the electroweakino sectors
are influenced by $\mu+\mu^\prime$ particularly when the combination that
is close to the higgsino mass is lesser than the masses of relevant gauginos
$M_1$  and $M_2$.  
We will discuss the phenomenological
implications on $\bsg$ and muon $g-2$ in this context.  
The diagrams that are significant in MSSM for $\bsg$ are the loops involving
$t-H^\pm$ and $\tilde t - \chonepm$. Agreement of the SM result of $\bsg$ with the 
experimental data demands delicate cancellation between the contributions of
the above loops. In MSSM, $\bsg$ increases with
$\tan\beta$\cite{Haisch:2012re}. Additionally, there are Next to Leading Order 
(NLO) contributions from squark-gluino loops due to corrections of 
bottom and top Yukawa couplings, particularly for large values of
$\tan\beta$. Here we will discuss the effects on $\bsg$ in relation to the
results of 
Ref.\cite{Ca:nh} that used unconstrained NH soft terms given at the weak scale
(NHSSM). The $\tilde t - \chonepm$ loop that is relevant for our study
does not have an appreciable effect from top squark L-R mixing since both
$A_t$ and $A_t'$ that start from vanishing values at the
messenger scale have only limited degree of evolutions.
This is in contrast to the 
larger possible values assumed by the above trilinear parameters as analyzed in
Ref.\cite{Ca:nh}.
Fig.\ref{fig:b_sg} shows the variation of $\bsg$ with $\mu^{\prime}_0$ 
in NHmGMSB for a given $\Lambda$ and $M_{mess}$. It shows some sharp
increase in $\bsg$ for a given zone of $\mu^{\prime}_0$, particularly for
a large $\tan\beta=40$. It is found
that this is indeed the region when $\mu+\mu^\prime$ or
the lighter chargino mass becomes small.
The effect was also seen earlier\cite{Solmaz2009,Un:2014afa2}. 
The corresponding mGMSB values of the same quantity 
are also shown where $\mchonepm$ is not small.
Apart from the cases with small $\mchonepm$,
$\bsg$ does not impose any serious
constraint on NHmGMSB parameter space even for a large value
$\tan\beta$ because of the smallness of $A_t$ and $A_t'$ and large values of
top-squark masses so as to satisfy the Higgs mass data. 
This is unlike NHSSM that recovers a large amount of 
parameter space discarded by $\bsg$ in MSSM for a large value of
$\tan\beta$\cite{Ca:nh}.

\begin{figure}[!htbp]
  \begin{center}
  \includegraphics[width=0.5\textwidth,angle=0]{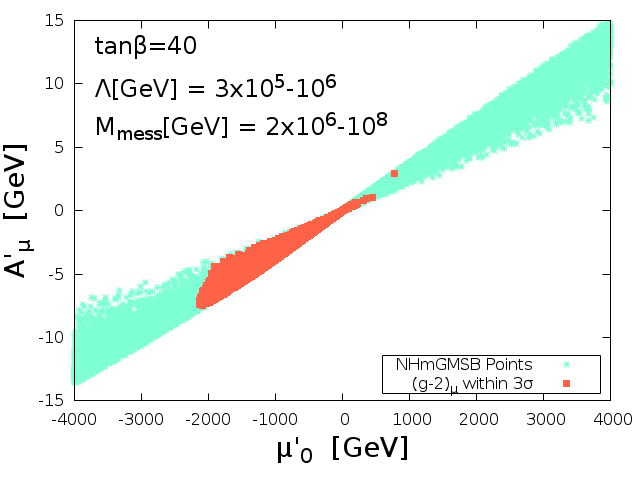}
 \caption{Scatter plot of $\mu^{\prime}_0$ vs $A_\mu'$ where $\Lambda$ and
   $M_{\rm mess}$ are varied according to the ranges mentioned in
   Eqs.\ref{scan_range}.
    The green points satisfy all the relevant constraints of 
Eq.\ref{experimental_bounds} while the orange points correspond to a region 
that satisfy the muon $g-2$ constraint at $3\sigma$ level.}
 \label{fig:mup_Amup}
 \end{center}
\end{figure}

We will now discuss the SUSY contributions to $\gmin2$,
namely $\amususy$~\footnote{$a_\mu=\frac{1}{2}{(g-2)}_\mu$} 
in the context 
of NHmGMSB against the results of the NHSSM analysis made in Ref.\cite{Ca:nh}.
At the one-loop level, $\amususy$ involves contributions from
${\tilde \chi}_i^\pm -
{\tilde \nu}_\mu $ and ${\tilde \chi}_i^0 -{\tilde \mu}$ loops\cite{muong1}.
In the NHSSM analysis of Ref.\cite{Ca:nh}, because of a strong L-R
mixing via $A_\mu'$, it is the ${\tilde \chi}_i^0 -{\tilde \mu}$
loop that contributes a significantly large amount to $\amususy$.
$A_\mu'$ can be as small as 100~GeV or even 50~GeV to show
very prominent effects. The largeness of the effect in NHSSM comes
from an enhancement of $A_\mu'$ via $\tan\beta$ and the diagram containing
bino-${\tilde \mu}_{L,R}$ in the loop that contributes to
$\amususy$\cite{hagiwara_muong,Endo_muong_set1}.
In NHmGMSB, $A_\mu'$ that starts from a vanishing value at $M_{\rm mess}$
has only a limited degree of 
evolution\cite{Ross:2016pml} even with a large $\mu^{\prime}_0$.   
As a result, with an inadequate level of enhancement of $A_\mu'$,
we do not expect any 
large amount of contributions from the bino-smuon loops due to L-R mixing,
unlike the NHSSM results of Ref.\cite{Ca:nh}.
Fig.\ref{fig:mup_Amup} shows a scatter plot of
$\mu^{\prime}_0$ vs $A_\mu'$ 
where $\Lambda$ and $M_{\rm mess}$ are varied within the shown ranges of
Eq.\ref{scan_range}. 
The green points are those that satisfy all the relevant constraints of 
Eq.\ref{experimental_bounds} while the orange points correspond to a region 
that satisfy the muon $g-2$ constraint at $3\sigma$ level. The $3\sigma$
region spreads more toward the negative zone of $\mu^{\prime}_0$. We should remind
that $\mu$ is obtained via REWSB with a chosen positive sign,
whereas $\mu^{\prime}_0$ assumes both the signs. Thus, a larger
$\amususy$ via a lighter higgsino mass can be obtained only in the negative
direction of $\mu_0^\prime$.   

\begin{figure}[!htbp]
 \begin{center}
  \includegraphics[width=0.5\textwidth,angle=0]{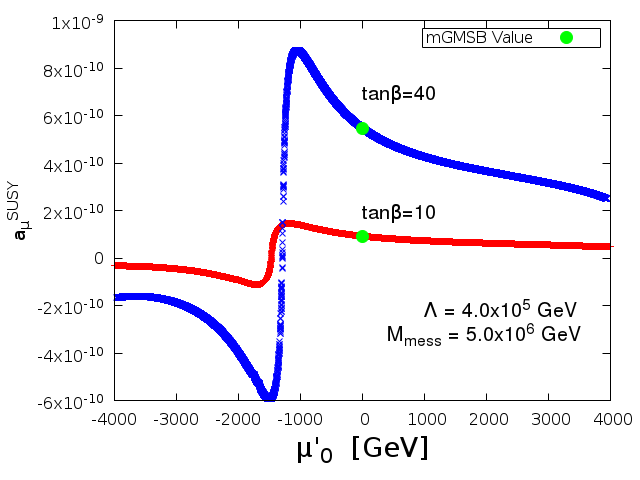}
  
  \caption{Plot of $\amususy$ vs $\mu^{\prime}_0$ for fixed $\Lambda$ and $M_{\rm mess}$ for
    $\tan\beta=10$ and $40$. The corresponding mGMSB values of the
    same quantity 
    are also shown by considering vanishing $\mu^{\prime}_0$.}
 \label{fig:g-2}
 \end{center}
\end{figure}

The fact that the large $A_\mu'$ regions are not associated with larger values 
of $\amususy$, rather the region with smaller $\mu'_0$ satisfy the 3$\sigma$\footnote{We consider Ref.\cite{Davier:2017zfy} for the following: 
  $a_\mu^{\rm SM}=   11659182.3 \pm 4.3 ~(\times 10^{-10}) $ and
  $a_\mu^{\rm exp}=11659209.1 \pm 6.3 ~(\times 10^{-10})$. This 
leads to a discrepancy of $26.8 \pm 7.6 ~(\times 10^{-10}) ~~(3.5\sigma)$ .}
level of the constraint predominantly for a negative region of $\mu^\prime$
indicates the domination of charged and neutral higgsinos in the
loop diagrams\cite{hagiwara_muong}.
This is indeed displayed in Fig.\ref{fig:g-2} that shows a 
plot of $\mu^{\prime}_0$ vs $\amususy$ for a given $\Lambda$ and
$M_{\rm mess}$. 
The figure shows a sharp rise in $\amususy$ for a small zone of $\mu^{\prime}_0$
that also increases with $\tan\beta$.
It turns out that with the value of $\mu$ 
obtained from REWSB, and $\mu+\mu^\prime$ being the tree level higgsino mass,
this rise corresponds to the light higgsino zone~\footnote{$\mu^{\prime}$ has a limited degree of evolution from $M_{mess}$ to the electroweak scale.}. 
The specific contributions to $\amususy$ arise from
diagrams\cite{hagiwara_muong} like wino/higgsino-sneutrino and
wino/higgsino-${\tilde \mu}_L$, bino/higgsino-${\tilde \mu}_L$ or 
bino/higgsino-${\tilde \mu}_R$.
There may also be a significant degree of cancellations between
the above diagrams. The 3$\sigma$ level of $\amususy$ region for $\mu^{\prime}_0$
as shown in Fig.\ref{fig:mup_Amup} comes out to be around $-1500$~GeV to
$200$~GeV.

\subsection{Higgsino like NLSP decays}
The interaction Lagrangian of gravitino $\psi_\mu$ with other sparticles and
SM particles is given by
$\mathcal L =\sum_{\alpha=1}^3 \mathcal{L}^{(\alpha)}$ where
$\alpha$ stands for a given gauge group out of ${SU(3)}_C \times {SU(2)}_L \times {U(1)}_Y$ and $\mathcal{L}^{(\alpha)}$ being given as\cite{Pradler:2007ne},    
\begin{equation*}
\label{gravitino_interaction}
 \mathcal{L}^{(\alpha)} = -\frac{i}{\sqrt{2}M_p}\large [{\cal D}_\mu^{(\alpha)} \phi^{*i}\bar\psi_\nu\gamma^\mu\gamma^\nu\chi^i_L - {\cal D}_\mu^{(\alpha)} \phi^{i}\bar\chi^i_L\gamma^\nu\gamma^\mu\psi_\nu\large]
 -\frac{i}{8M_p}\bar\psi_\mu [\gamma^\rho,\gamma^\sigma] \gamma^\mu\lambda^{(\alpha)a} F_{\rho\sigma}^{(\alpha)a}.  
\end{equation*}
The covariant derivatives ${\cal D}_\mu^{(\alpha)}$ and $F_{\rho\sigma}^{(\alpha)a}$
are appropriately defined depending on the gauge group denoted by $\alpha$ and
the generator index $a$\cite{Pradler:2007ne}.

\noindent
The resulting decay widths $\Gamma ( \widetilde{\chi}^0_1 \to \widetilde{G}\ + Z)$ and 
$\Gamma ( \widetilde{\chi}^0_1 \to \widetilde{G} + h )$ 
\cite{Dimopoulos:1996yq, Giudice:1998bp,Covi:2009bk,Ambrosanio:1996jn,Ambrosanio:1999iu} are given as below. 
This is keeping in mind that $\lspone$, the NLSP candidate, is chosen to be
higgsino dominated in nature and the decays into heavier Higgs bosons are
kinematically disallowed. We also note that a higgsino 
type of NLSP couples only to the 
longitudinal $Z$ component. In realistic scenarios with small amount of
gaugino mixing in NLSP, which we will though ignore here,   
$\Gamma ( \widetilde{\chi}^0_1 \to \widetilde{G}\, \gamma)$ can be quite
important. However, we will work only with an almost pure higgsino. For the above two decays we have,
\begin{equation}
\label{decay_width_Z}
\Gamma ( \widetilde{\chi}^0_1 \to \widetilde{G}\, Z)  \simeq\frac{m_{\widetilde{\chi}^0_1}^5}{96 \pi\, m_{\tilde G}^2 M_p^2} 
 \left|-N_{13} \cos\beta + N_{14} \sin\beta \right|^2\left(1-\frac{m_{Z}^2}{m_{\widetilde{\chi}^0_1}^2}  \right)^4,
\end{equation}

\begin{equation}
\label{decay_width_h}
\Gamma ( \widetilde{\chi}^0_1 \to \widetilde{G}\, h ) \simeq\frac{m_{\widetilde{\chi}^0_1}^5}{96 \pi\, m_{\tilde G}^2 M_p^2} 
\left|-N_{13} \sin\alpha + N_{14}\cos\alpha \right|^2
\left(1-\frac{m_{h}^2}{m_{\widetilde{\chi}^0_1}^2}  \right)^4, 
\end{equation}
where $N_{13}$ and $N_{14}$ are the higgsino related components of
the LSP in the $\tilde B, \tilde W, \widetilde{H_d^0}, \widetilde{H_u^0}$ basis
of the neutralino diagonalizing matrix $N_{ij}$\cite{SUSYbook1}.
The gravitino mass is given by, 
\begin{equation*}
\label{grav_mass}
 m_{\tilde G} = \frac{\Lambda M_{mess}}{\sqrt{3}M_P}.
\end{equation*}

\begin{figure}[!htbp]
 \begin{center}
   \subfigure[]{%
     \includegraphics[width=0.47\textwidth,angle=0]{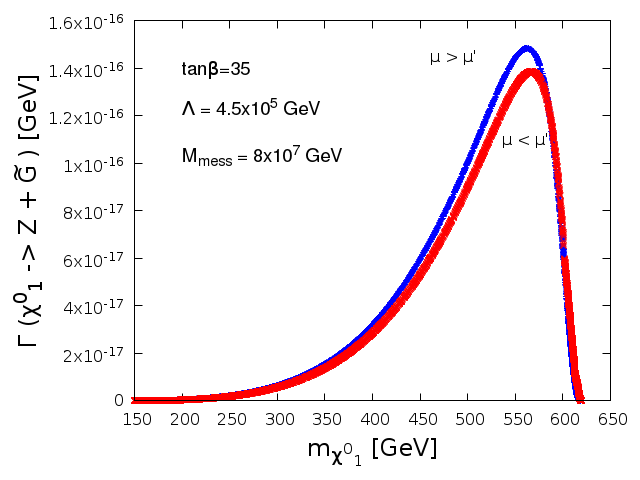}
     \label{fig:decayA}
  } 
  \subfigure[]{
    \includegraphics[width=0.47\textwidth,angle=0]{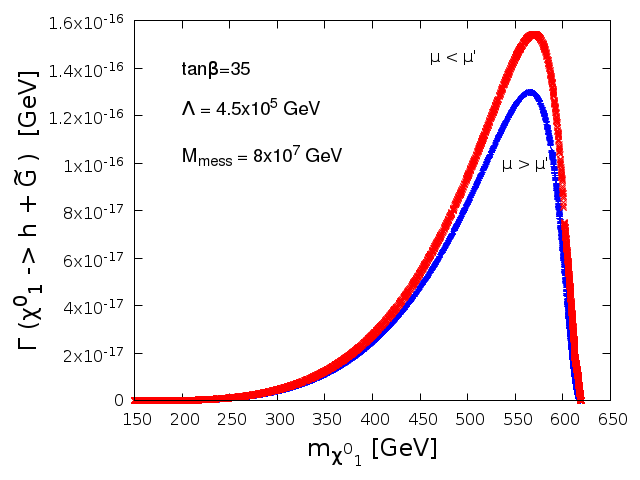}
    \label{fig:decayB}
  }
  \caption{Plot of decay width vs. NLSP mass  
    for $\widetilde{\chi}^0_1 \to \widetilde{G}\ + Z$ and
    $\widetilde{\chi}^0_1 \to \widetilde{G}\ + h$ channel for $\mu>0$ when
    $\mu^\prime$ is scanned over a range (Eq.\ref{scan_range})
    to probe the higgsino NLSP zone. 
    The blue and red points refer to $\mu > |\mu^{\prime}|$ and
    $\mu < |\mu^{\prime}|$ respectively.}
\label{fig:decay}
 \end{center}
\end{figure}

For our choice of parameter space $m_{\tilde G}$ is of order of few keV. This 
allows us to neglect it in the expression of the  
phase space factor compared to other
masses involved in the calculation of decay widths 
in Eqs:\ref{decay_width_Z},\ref{decay_width_h}.
 The total decay width of a higgsino type of NLSP is given by,
\begin{equation}
\Gamma^{\rm tot} \equiv \Gamma^{\rm tot}_{\rm NLSP} \simeq \Gamma ( \widetilde{\chi}^0_1 \to \widetilde{G}\, Z)
+ \Gamma ( \widetilde{\chi}^0_1 \to \widetilde{G}\, h).
\label{Gamma_tot_expression} 
\end{equation}
$\Gamma^{\rm tot}$ would be strongly influenced by 
$\Lambda$ and $M_{mess}$ since $\Gamma^{\rm tot} \propto \frac{1}{(\Lambda M_{mess})^2}$. 
$\Gamma^{\rm tot}$ is also influenced by $\mu^{\prime}_0$ through the higgsino NLSP 
mass as well as the effect of higgsino mixing in $N_{ij}$s. 
We explore the decay widths of Eqs.\ref{decay_width_Z} and
\ref{decay_width_h} against the NLSP mass in Fig.\ref{fig:decayA} and \ref{fig:decayB} respectively. The fixed
values of the chosen parameters are sign$(\mu)=1$, $\tan\beta=35$,
$\Lambda=4.5 \times 10^5$~GeV and $M_{mess}=8 \times 10^7$~GeV.
$\mu^{\prime}_0$ is scanned over a
range $-4$~TeV to $4$~TeV.
Both the decay widths increase with NLSP mass as long as the
  NLSP is substantially a higgsino. This is so until $\mlspone$ is
  about 550~GeV.
Once, the NLSP is found to be an admixture of a higgsino and bino
or essentially a bino
each of the decay widths is bound to decrease rapidly. This happens 
after it attains a peak similar to what is apparent in  
Fig.\ref{fig:decayA} or \ref{fig:decayB}. 
In short of other relevant decay modes,
Eq.\ref{Gamma_tot_expression} would then no longer be valid. 
Regarding the composition of the NLSP and its effect on $\Gamma ( \widetilde{\chi}^0_1 \to \widetilde{G}\, Z)
$ and $\Gamma ( \widetilde{\chi}^0_1 \to \widetilde{G}\, h)$ we would like to point out that
both $N_{13}$ and $N_{14}$ change signs over the scanned region of parameter space. However,
$N_{14}$ plays a dominant role for both types of decays unless $\tan\beta$ is very small.
In Eq.\ref{decay_width_Z} the $N_{13}$ contribution is
very subdominant since $\cos\beta$ is small.
A similar subdominant role for $N_{13}$ is also true in
Eq.\ref{decay_width_h} since we have a decoupling Higgs boson
scenario where $\sin\alpha$ becomes small\cite{djouadi}.
The above causes both $\Gamma ( \widetilde{\chi}^0_1 \to \widetilde{G}\, Z)
$ and $\Gamma ( \widetilde{\chi}^0_1 \to \widetilde{G}\, h)$ to peak at
similar values of the NLSP mass.  
Apart from the above,  because of the possibility of both the signs
for $N_{13}$ and $N_{14}$,  
Figs.\ref{fig:decayA} and \ref{fig:decayB} show that each of the 
decay widths has two branches or in other words is 
a double-valued function of $\mlspone$ depending 
on the two cases $\mu > |\mu^{\prime}|$ and $\mu < |\mu^{\prime}|$. 
The heights of the decay widths of the two branches for 
each of $\Gamma ( \widetilde{\chi}^0_1 \to \widetilde{G}\, Z)
$ and $\Gamma ( \widetilde{\chi}^0_1 \to \widetilde{G}\, h)$  
differ because of a varying degree of radiative corrections to
the NLSP mass. 
Considering $\mlspone \simeq 575$~GeV (near the peaks)
the approximate sum of the two 
decay widths is around $2.8 \times 10^{-16}$~GeV for the given parameter point.
%an approximate maximum height of $6\times 10^{-7}$~GeV for the considered param%eter point,
This leads to $1/\Gamma^{\rm tot} \simeq 2 \times 10^{-9} ~{\rm sec}
\simeq 70$~cm. In general,  
the mean decay length of $\chi_{1}^{0}$ as NLSP 
with energy E in the laboratory frame is given 
by\cite{Giudice:1998bp, Ambrosanio:1996jn,Ambrosanio:1999iu},  
\begin{equation}
\label{decay_length}
d ~ \simeq ~ (E^2/m^2_{\chi^0_1}-1)^{1/2}/\Gamma^{\rm tot}. 
\end{equation}

\begin{figure}[!htbp]
 \begin{center}
   \subfigure[]{%
     \includegraphics[width=0.47\textwidth,angle=0]{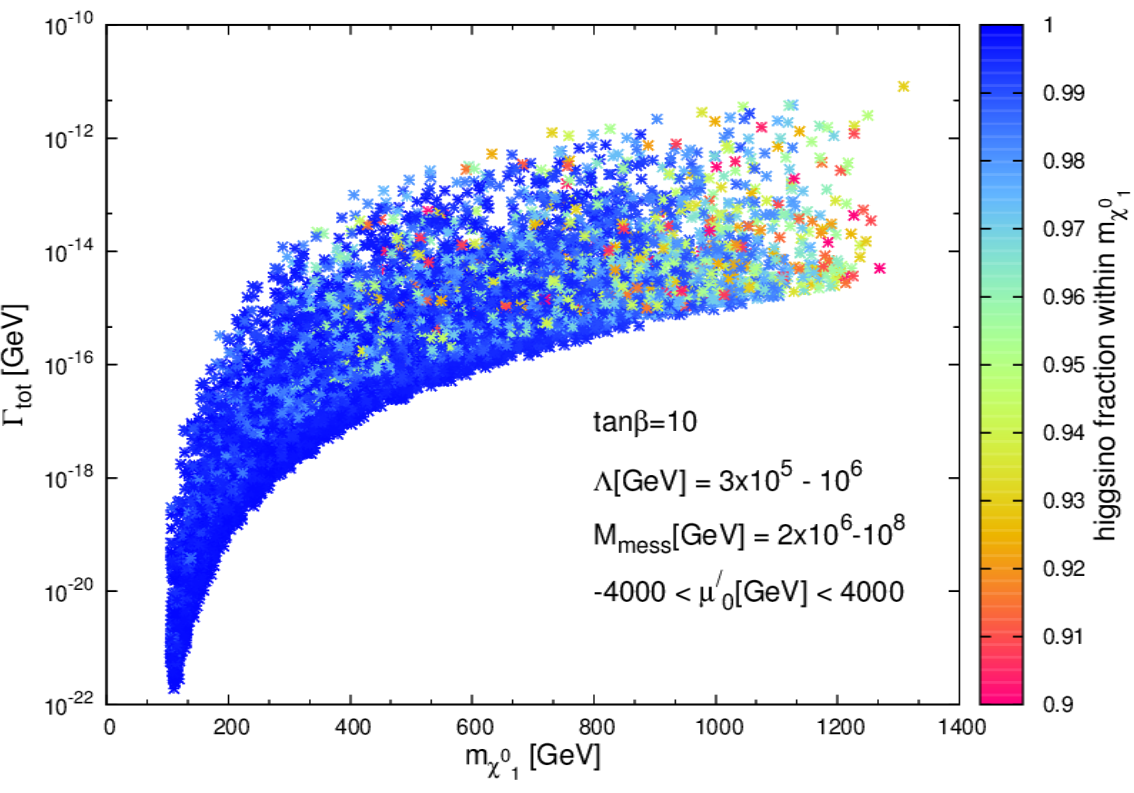}
     \label{fig:scanneddecaywidthA}
  } 
  \subfigure[]{
    \includegraphics[width=0.47\textwidth,angle=0]{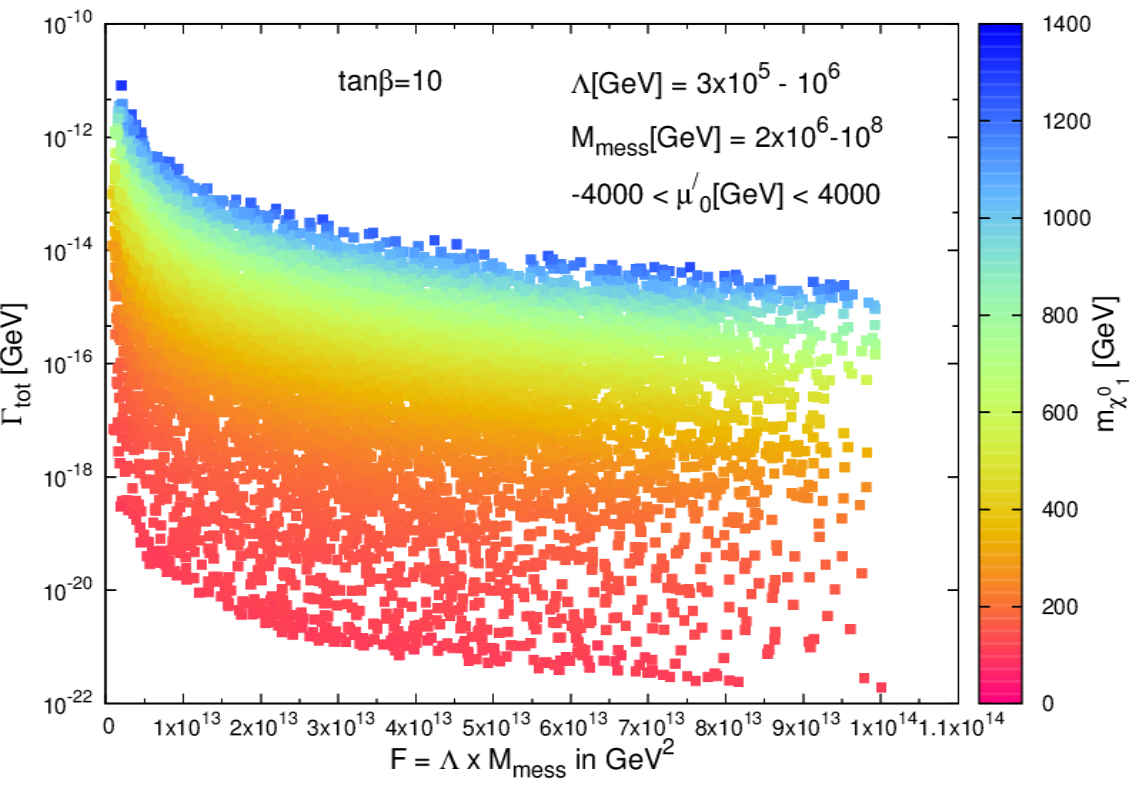}
    \label{fig:scanneddecaywidthB}
  }
  \caption{(a) Scatter plot of decay width $\Gamma^{\rm tot}$ vs.   
    $\mlspone$ for a higgsino dominated NLSP over the scanned parameter
    region of Eq.\ref{scan_range}. The higgsino fraction is shown in graded
    color with a reference color bar on the right.
    (b) Similar scatter plot in the plane of $\Gamma^{\rm tot}$ vs. $F$ where
    the NLSP mass is shown in a graded color with a reference color bar on the
    right.
}
\label{fig:scanneddecaywidth}
 \end{center}
\end{figure}

Fig.\ref{fig:scanneddecaywidthA} shows a scatter plot of decay width
$\Gamma^{\rm tot}$ vs. $\mlspone$ for a higgsino dominated NLSP over the scanned parameter
    region of Eq.\ref{scan_range}. The higgsino fraction is shown in graded
    color with a reference color bar on the right. Only highly higgsino
    dominated NLSP region is considered. 
    Fig.\ref{fig:scanneddecaywidthB} shows a similar scatter plot in the
    plane of $\Gamma^{\rm tot}$ vs. $F(=\Lambda M_{\rm mess})$ where
    the NLSP mass is shown in a graded color with a reference color bar on the
    right.  
    The range of variation of $\Gamma^{\rm tot}$ is from
    $10^{-22}$ to $10^{-12}$~GeV implying $1/\Gamma^{\rm tot}$ to be within
    $\simeq 10^{-3}$~sec to $\simeq 10^{-13}$~sec or $\simeq 1000$~km to
    $0.1$~mm respectively. The decay lengths when computed
    in the laboratory frame would point out a long
    range of values indicating decays occurring both within and outside the detector .
    Collider studies of probing the higgsino NLSP decays for suitable
    values of $\Lambda M_{\rm mess}$ may be performed 
    similar to the analyses made in  
    Refs.\cite{Bobrovskyi_2011-12}. This is however beyond the scope of the
    present work. 

%\clearpage
Finally, we present two representative points A and B in Table~\ref{bptable}
for the spectra of NHmGMSB to demonstrate
the degree of evolution of the parameters connected with the NH terms while
choosing the lighter chargino and the NLSP to be higgsino dominated in nature.
This is generally unavailable in mGMSB where the NLSP as the lightest
neutralino it is typically bino dominated in its composition.
The spectra is generally heavy because of the requirement of satisfying
the Higgs boson mass limit. We remind that the effect of the NH terms can be
increased significantly and the 
spectra may be lighter while that would also satisfy the muon $g-2$ data,  
if we go beyond the mGMSB 
setup. This may be possible if the trilinear couplings for NH terms
may have non-vanishing values at the messenger scale. 
%%%%%%
 \begin{center}
\begin{table}[!htbp]
 \caption{Representative Points for NHmGMSB: All the dimensional parameters are in GeV.}
\label{bptable}  
 \centering

  \begin{tabular}{|c||c|c|}
  
  \hline\hline
  
  Parameters  &  A &  B   \\ [0.5ex]
\hline
  $\Lambda$ & $3.65\times10^5$ & $3.16\times10^5$ \\
  $M_{mess}$ & $9.742\times10^6$ & $8.073\times10^6$\\
  $\tan\beta$ & 10 & 40 \\
  $A_0^{\prime}$ & 0 & 0\\
  $\mu_0^{\prime}$ & -1898 & -1144\\
  
\hline\hline

$A_t$ & -787 & -686 \\
$A_b$ & -136 & -430 \\
$A_\tau$ & -14 & -38 \\

$A_t^{\prime}$ & -210 & -147 \\
$A_b^{\prime}$ & -55 & -121 \\
$A_\tau^{\prime}$ & -23 & -57 \\  
$A_\mu^{\prime}$ & -1.4 & -3.4 \\

$m_h$ & 122.1 & 122.3 \\
$m_H,m_{H^{\pm}},m_A$ & {2047,2047,2047} & {1425,1425,1425}\\
$m_{\tilde t_{1,2}}$ & {3090,3458} & {2651,2949}  \\
$m_{\tilde b_{1,2}}$ & {3357,3453} & {2841,2946}  \\
$m_{\tilde \tau_{1,2}}$ & {695,1315} & {566,594} \\
$m_{\tilde \chi_{1,2}^0}$ & {432,451} & {202,212} \\
$m_{\tilde \chi_{1,2}^{\pm}}$ & {446,981} & {210,846} \\
$m_{\tilde g}$ & 2636 & 2311 \\
\hline
NLSP Composition & $\tilde{\chi}^0_1 \approx 86 \% \tilde H$ like & $\chi^0_1 \approx 98 \% \tilde H$ like  \\

$BR(B \rightarrow X_s +\gamma)$ & $3.22\times10^{-4}$ & $3.21\times10^{-4}$ \\
$BR(B_s \rightarrow \mu^+ \mu^-)$ & $3.27\times10^{-9}$ & $3.28\times10^{-9}$\\
$a_{\mu}^{SUSY}$ & $1.027 \times 10^{-10}$ & $7.88 \times 10^{-10}$ \\
\hline\hline 
\end{tabular}
 \end{table}
 \end{center}

\clearpage
\section{Conclusion}
\label{sec:conclusion}
It was seen that SUSY models may  
include non-holomorphic terms like $\phi^2 \phi^*$ and
$\psi \psi$ that can be characterized as soft SUSY breaking in nature
in the absence of a gauge singlet field. 
The broad applicability of the terms in various possible SUSY models 
makes such inclusion very important.     
In particular, the above terms may relax stringency to accommodate various
phenomenological data in MSSM. 
These would be additional interactions than the 
usual soft SUSY breaking terms like that for the scalar masses
(non-holomorphic) and trilinear and bilinear interactions that are
holomorphic in nature. 
There have been studies 
on non-holomorphic MSSM referred as NHSSM that included such terms.
The specific areas of impact
that NHSSM makes are in the phenomenologies involving i) L-R mixing of
squarks or sleptons via trilinear non-holomorphic terms containing
the conjugate higgs fields, ii) higgsino mass soft SUSY breaking 
non-holomorphic term that results into higgsino components of electroweakinos
to have parts coming from both superpotential as well as
soft breaking origins and iii) 
the tree level electroweak fine-tuning to have no essential 
correlation with the higgsino mass, unlike MSSM.  
The authors of Refs.\cite{Ca:nh,Beuria:2017gtf} 
performed phenomenological MSSM (pMSSM) type of analysis 
where all the soft parameters including the NH ones 
are provided at the weak scale. These were 
away from previous works that used 
mostly CMSSM inspired setup where in addition to the usual
CMSSM inputs corresponding to the gauge coupling unification scale, the
NH soft parameters were either given a) at the
unification scale or b) at the electroweak scale.
It was however shown that in the absence of a gauge singlet superfield,
a hidden sector based  
F-type SUSY breaking scenario with two chiral superfields would 
lead to such NH soft terms\cite{Martin:nh}. The terms
arise out of D-term contributions. 
On the other hand, supergravity scenarios with high scale SUSY breaking could 
include such NH terms with mass 
scale suppression nearing the Planck scale\cite{Martin:nh}.
Irrespective of the above, there have been analyses
that worked out the phenomenologies of including such soft terms in a
CMSSM setup or extensions like non-universal gaugino or non-universal
Higgs scalar scenarios.

%\noindent
In this work, we first investigate the NH soft SUSY breaking
models in a low scale SUSY breaking context. We choose to perform the analysis
within a backbone provided by the minimal Gauge mediated SUSY
Breaking (mGMSB) framework.  
We however assume that the higgsino mixing soft term parameter
$\mu^\prime$ to have a SUSY breaking origin away from mGMSB.  
Thus, by having an entirely independent SUSY breaking origin for the
higgsino mass soft term we essentially
overcome any issue of the reparametrization invariance involving
unrelated quantities as mentioned in the text.   
We focus on the degree of influence of the 
NH terms in an mGMSB setup (NHmGMSB) in comparison with the NHSSM work mentioned earlier. We note that the NH terms may affect various SUSY parameters
including the Higgs scalar mass parameters through the RGEs. Thus, $\mu$, 
as obtained via REWSB may be non-negligibly modified once the NH terms are
included.

We show that the level of the L-R mixing of top-squarks  
can potentially reduce the $\mstopone$ to a reasonably 
large extent, in turn increasing the radiative corrections to Higgs boson
mass. Thus, a smaller than required
$|A_t|$ value can do the job of satisfying
the lower bound of $m_h$ compared to the mGMSB case since 
a compensating contribution may come from appropriate NH couplings. 
Of course, the 
enhancement of radiative corrections to the higgs boson mass is lesser when 
we compare our result with that of NHSSM. This is simply due to the limited
degree of RG evolution of $A_t$ and $A_t'$ from $M_{\rm mess}$ down 
to the electroweak scale starting from their vanishing values at
the higher scale. Regarding the effects on the 
electroweakino sector, we observe that the high 
level of enhancement (via a factor of $\tan\beta$) of the SUSY 
contributions to muon $g-2$ as that happens in NHSSM via $A_\mu^\prime$ through
the L-R mixing contributions of smuons is absent in NHmGMSB. This is  because
$A_\mu'$ does not become sufficiently large at the weak scale.
The smallness of $A_\mu'$ can, however, be avoided in non-minimal GMSB cases
where the NH trilinear coupling $A_0'$ need not be vanishing at the messenger scale.
Simultaneously, the higgs boson mass $m_h$ would receive a larger amount of
radiative correction through enhanced $A_t'$.  A larger 
$\amususy$ comes from the limited zone of light higgsino mass in NHmGMSB.
The same 
light higgsino mass zone may enhance the $\bsg$ contribution via the
chargino loops. However, the constraint from $\bsg$ is well satisfied at
the 2$\sigma$ level over the 
relevant parameter space of NHmGMSB simply because of generally large 
top-squark masses. This in turn arises from the requirement
of satisfying the Higgs mass bound. We further estimate the decay of
the lightest neutralino as NLSP into gravitino and $Z$-boson
or $h$-boson while considering only a higgsino dominated NLSP.
This is in contrast to a typical mGMSB scenario 
where $\lspone$ is bino-dominated in its composition.
We probed the entire parameter space of
NHmGMSB to find a large degree of variation of decay lifetimes that may
correspond to a decay length of less than a millimeter to hundreds of
kilometer in the rest frame of the NLSP. 
Relevant collider analyses may be made by finding the length 
in the laboratory frame of the detector. Depending 
on whether the NLSP decay is happening within or outside the detector one can
probe NHmGMSB for a higgsino type of NLSP. Finally, we have
presented two representative points for the spectra of NHmGMSB to
demonstrate the extent of RG evolutions parameters related to 
the NH soft terms. The two 
spectra are on the heavier side so as to accommodate the Higgs mass data.
However, a non-minimal GMSB scenario that allows having non-vanishing
trilinear parameters $A_0$ and $A_0'$ at the messenger scale would show
a very significant effect on the scalar sector including the Higgs mass apart
from potentially providing an enhanced $\amususy$ via a
large L-R mixing of the smuons.
This is beyond the focus of the present analysis.
Finally, considering the generic nature of the NH soft terms,
it is important to explore their effect in varying
SUSY breaking scenarios and scales. It can also be significant in respect of global analyses
of various SUSY models.

\noindent
    {\bf Acknowledgments:} UC would like to thank Anirban Kundu and Sourov Roy
    for a brief discussion. UC and SM are grateful to Abhishek
    Dey for many helpful suggestions.   
\clearpage
%%%%%%%%%%%%%%%%%%%%%%%%%%%%%%%%%%%%%%%%%%%%%55
%\newpage

\end{document}